\pdfoutput=1
\documentclass[12pt,a4]{article}
\usepackage{relsize}
\usepackage{array}
\usepackage{a4wide}
\usepackage[hidelinks]{hyperref}
\usepackage{latexsym}
\usepackage{cite}
\usepackage{mathrsfs}
\usepackage{amssymb}
\usepackage{amsmath}
\usepackage{graphicx}
\usepackage{enumerate}
\usepackage[usenames,dvipsnames]{xcolor}
\usepackage{todonotes}

\newcommand{\p}{\partial}

\newcommand{\1}{{\rm 1\hskip-0.25em I}}
\newcommand{\be}{\begin{equation}\label}
\newcommand{\ee}{\end{equation}}
\newcommand{\bea}{\begin{eqnarray}\label}
\newcommand{\eea}{\end{eqnarray}}

\newcommand{\lambdat}{\tilde{\lambda}}
\newcommand{\xit}{\tilde{\xi}}

\makeatletter
\newcommand*{\textoverline}[1]{$\overline{\hbox{#1}}\m@th$}
\newcommand*\bigcdot{\mathpalette\bigcdot@{.65}}
\newcommand*\bigcdot@[2]{\mathbin{\vcenter{\hbox{\scalebox{#2}{$\m@th#1\bullet$}}}}}
\makeatother

\newcommand{\ang}[1]{\left\langle #1\right\rangle}
\newcommand{\squ}[1]{\left[ #1\right]}
\newcommand{\sbra}[1]{\lambdat_{#1}}
\newcommand{\sket}[1]{\lambdat_{#1}}
\newcommand{\bra}[1]{\lambda_{#1}}
\newcommand{\ket}[1]{\lambda_{#1}}

\newcommand{\vertexopleftdelta}{\;\delta^{2}\left(\tilde{\lambda}_{i}-t\tilde{\lambda}(\sigma)\right) e^{it\ang{ \tilde{\mu}(\sigma) i} }\;}
\newcommand{\vertexoprightdelta}{\;\delta^{2}\left(\lambda_{i}-t\lambda(\sigma)\right)e^{it\squ{\mu(\sigma)i}}\;}
\newcommand{\sigmeasure}{\frac{d^{2\times n}\sigma}{GL(2)}}

\newcommand{\pd}[1]{\frac{\partial}{\partial #1}}
\newcommand{\pdd}[2]{\frac{\partial #1}{\partial #2}}

\date{}
\begin{document}

\title{New Worldsheet Formulae for Conformal Supergravity Amplitudes}
\author{Joseph A. Farrow and Arthur E. Lipstein \vspace{7pt}\\ \normalsize \textit{
Department of Mathematical Sciences}\\\normalsize\textit{Durham University, Durham, DH1 3LE, United Kingdom}}
\maketitle
\begin{abstract}
We use 4d ambitwistor string theory to derive new worldsheet formulae for tree-level conformal supergravity amplitudes supported on refined scattering equations. Unlike the worldsheet formulae for super-Yang-Mills or supergravity, the scattering equations for conformal supergravity are not in general refined by MHV degree. Nevertheless, we obtain a concise worldsheet formula for any number of scalars and gravitons which we lift to a manifestly supersymmetric formula using four types of vertex operators. The theory also contains states with non-plane wave boundary conditions and we show that the corresponding amplitudes can be obtained from plane-wave amplitudes by applying momentum derivatives. Such derivatives are subtle to define since the formulae are intrinsically four-dimensional and on-shell, so we develop a method for computing momentum derivatives of spinor variables.
\end{abstract}

\pagebreak
\tableofcontents

\section{Introduction}

Starting with the discovery of the Parke-Taylor formula for tree-level gluon amplitudes \cite{Parke:1986gb}, the study of scattering amplitudes has suggested intruiging new ways of formulating quantum field theory. Building on the work of Nair \cite{Nair:1988bq}, Berkovits and Witten \cite{Witten:2003nn,Berkovits:2004hg} proposed a worldsheet model for $\mathcal{N}=4$ super-Yang-Mills (SYM) known as twistor string theory, whose correlation functions give rise to an elegant formula for tree-level amplitudes in terms of integrals over curves in twistor space \cite{Roiban:2004yf}. It turns out however that twistor string theory also contains conformal supergravity (CSGR) in its spectrum \cite{Berkovits:2004jj}, so it is difficult to extend this formula to loop-level. These ideas were then extended by Skinner to $\mathcal{N}=8$ supergravity (SUGRA) \cite{Skinner:2013xp} following the discovery of a gravitational analogue of the Park-Taylor formula by Hodges \cite{Hodges:2012ym}. This framework was subsequently extended to a broad range of theories after Cachazo, He, and Yuan (CHY) proposed worldsheet formulae for tree-level scattering amplitudes of non-supersymmetric gauge and gravitational theories in any dimension\cite{Cachazo:2013hca}. These formulae are supported on solutions to the scattering equations, which were previously discovered in the context of ordinary string theory \cite{Fairlie:1972zz,Gross:1987kza}. The worldsheet theory underlying the CHY formulae was constructed by Mason and Skinner and is called ambitwistor string theory\cite{Mason:2013sva}. In the case of 10d supergravity, this model is critical and can be extended to loop-level \cite{Adamo:2013tsa,Geyer:2015bja}. 

Using a model known as 4d ambitwistor string theory, it is possible to obtain  manifestly supersymmetric formulae for tree-level scattering amplitudes in 4d $\mathcal{N}=4$ SYM and $\mathcal{N}=8$ SUGRA supported on refined scattering equations \cite{Geyer:2014fka,Bandos:2014lja}. In this approach the scattering equations are split into two sets, which we refer to as left and right. In general, the number of states in the left set is related to the Grassmann degree of the superamplitude. Furthermore, in the case of SYM and SUGRA, it is also tied to the MHV degree of the amplitude, which is related to the number of negative helicty supermultiplets being scattered. In particular, an N$^{k-2}$MHV amplitude describes the scattering $k$ negative helicity multiplets. These formulae are simpler than twistor string formulae in that they do not contain integrals over moduli of curves in twistor space, and are also simpler than the CHY formulae in general dimensions in that the solutions to the refined scattering equations are split into different MHV sectors\footnote{In particular, the number of solutions to the refined scattering equations is given by the Eulerian number $\left\langle\begin{smallmatrix} n-3\\k-2 \end{smallmatrix}  \right\rangle$ for an $n$-point N$^{k-2}$MHV amplitude \cite{Spradlin:2009qr,Cachazo:2013iaa,Lipstein:2015rxa}. In contrast, the scattering equations in general dimensions have $(n-3)!$ solutions.}. Moreover, the formulae arising from 4d ambitwistor string theory are intimately related to Grassmannian integral formulae for $\mathcal{N}=4$ SYM and $\mathcal{N}=8$ SUGRA obtained using on-shell diagrams \cite{ArkaniHamed:2012nw,Heslop:2016plj,Herrmann:2016qea,Farrow:2017eol}. 

Like the Berkovits-Witten twistor string, the 4d ambitwistor string for $\mathcal{N}=4$ SYM contains CSGR in its spectrum, in particular a certain non-minimal version of $\mathcal{N}=4$ conformal supergravity (in the minimal version there is no coupling between the Weyl tensor and scalar fields of the model \cite{Bergshoeff:1980is}). Although CSGR is not unitary, it is nevertheless of theoretical interest for several reasons. For example, it is renormalizable and can be made UV finite if coupled to $\mathcal{N}=4$ SYM \cite{Fradkin:1983tg,Fradkin:1985am}. Furthermore, it is possible to obtain classical Einstein gravity with cosomological constant by imposing Neumann boundary conditions on conformal gravity \cite{Maldacena:2011mk}  (see also \cite{anderson,Anastasiou:2016jix}), and this approach was used to deduce twistor string formulae for scattering amplitudes of Einstein supergravity in flat space \cite{Adamo:2012nn,Adamo:2013tja}.  

Given the large amount of symmetry in CSGR, we expect its scattering amplitudes to have simple mathematical properties. The purpose of this paper is therefore to investigate this structure using 4d ambitwistor string theory. CSGR amplitudes were previously studied in \cite{Berkovits:2004jj,Ahn:2005es,Dolan:2008gc,Brodel:2009ep,Adamo:2012xe,Adamo:2018srx}. More recently, these amplitudes were shown to arise from taking the double-copy of super-Yang-Mills with a $(DF)^2$ gauge theory \cite{Johansson:2017srf}. An ambitwistor string description of the $(DF)^2$ theory was subsequently found in \cite{Azevedo:2017lkz} and used to deduce a CHY formula for conformal gravity amplitudes in general dimensions. 

In this paper, we use 4d ambitwistor string theory to derive compact new worldsheet formulae for CSGR amplitudes supported on refined scattering equations. In contrast to the worldsheet formulae for $\mathcal{N}=4$ SYM and $\mathcal{N}=8$ SUGRA, we find that the number of particles in the left set is not generally tied to the MHV degree. Nevertheless, the formulae we obtain are very simple, allowing us to generalise previous results in many ways. For example, we obtain a simple formula for scalar-graviton amplitudes with any number of particles in the left set. If only two particles are in the left set, this formula reduces to the one previously derived by Berkovits and Witten. More generally, the formula can be readily evaluated numerically and we match it against results obtained from Feynman diagrams and double copy techniques developed in \cite{Johansson:2017srf} up to 8 points with any number of particles in the left set. Moreover, we generalize this to a manifestly supersymmetric formula using four types of vetex operators which describe states in either the left or right set and either the positive or negative helicity graviton multiplet.

Since the equations of motion for conformal gravity are fourth order in derivatives, they also admit non-plane wave solutions of the form $A\cdot x e^{ik\cdot x}$. We find that the vertex operators for such states are very simple and give rise to scattering amplitudes which can be obtained by taking momentum derivatives of plane wave amplitudes, and are therefore well-defined at least in a distributional sense. Vertex operators for non-plane wave states were previously proposed in \cite{Dolan:2008gc}. Whereas previous vertex operators were defined only for $A^2=0$, our vertex operators are defined for any $A$ and appear to be more compact. Nevertheless, computing non-plane wave amplitudes using 4d ambitwistor string theory turns out to be subtle for several reasons. First of all, in order to compute worldsheet correlators we introduce source terms in the path integral leading to deformed scattering equations. The final result is then obtained by taking functional derivatives with respect to the sources and setting them to zero, which returns to the original scattering equations. Second of all, since the formulae are manifestly four-dimensional and on-shell, we develop a prescription for taking momentum derivatives of spinor variables.     

The structure of this paper is as follows. In section \ref{rev}, we review some facts about 4d ambitwistor string theory that will be relevant in this paper. In section \ref{pw}, we derive worldsheet formulae for scattering amplitudes of graviton multiplet states with plane-wave boundary conditions. In section \ref{npwse}, we generalize these formulae to non-plane wave states, and in section \ref{conc} we present our conclusions and future directions. There are also four appendices. In appendix \ref{brst}, we review the BRST quantization of 4d ambitwistor string theory and show that the vertex operators presented in this paper are BRST invariant. In appendix \ref{MHV}, we show that our worldsheet formula for scalar-graviton amplitudes reduces to the Berkovits-Witten result when there are only two particles in the left set. In appendix \ref{momentumderv}, we describe a method for taking momentum derivatives of on-shell variables. Finally, in appendix \ref{npwexamples}, we use this method to compute several examples of non-plane wave amplitudes.       

\section{Review} \label{rev}

In this section, we will review the 4d ambitwistor string theory describing $\mathcal{N}=4$ SYM and CSGR. The Lagrangian for this model  is given by
\begin{equation}
\mathcal{L}=\frac{1}{2}\left(Z \cdot \bar{\partial}W-W \cdot \bar{\partial}Z\right)+\mathcal{L}_{j}
\label{action}
\end{equation}
where the worldsheet coordinate is a complex number $\sigma$ with $\bar{\partial}=\partial_{\bar{\sigma}}$, the target space is supertwistor space
\[
Z^{A}=\left(\begin{array}{c}
\lambda_{\alpha}\\
\mu^{\dot{\alpha}}\\
\chi^{I}
\end{array}\right),\,\,\, W_{A}=\left(\begin{array}{c}
\tilde{\mu}^{\alpha}\\
\tilde{\lambda}_{\dot{\alpha}}\\
\tilde{\chi}_{I}
\end{array}\right),
\]
and $\mathcal{L}_{j}$ is the Lagrangian for a current algebra, the details of which will not be important.  Note that $\alpha,\dot{\alpha}=1,2$ are spinor indices and $I=1,2,3,4$ is an R-symmetry index, and the corresponding worldsheet fields are bosonic and fermionic, repsectively. Furthermore, recall that a 4d null momentum can be written in bispinor form as follows:
\[
k_{i}^{\alpha\dot{\alpha}}=\lambda_{i}^{\alpha}\tilde{\lambda}_{i}^{\dot{\alpha}} 
\]
where $i$ labels the particle number. The worldsheet fields $\lambda,\tilde{\lambda}$ can therefore be thought of as parametrizing an on-shell momentum, although their relation to the external spinors of the amplitude will be made precise later on. The scattering amplitudes will ultimately be expressed in terms of inner products $\left\langle ij\right\rangle =\lambda_{i}^{\alpha}\lambda_{j}^{\beta}\epsilon_{{\alpha}{\beta}}$ and $\left[ij\right]=\tilde{\lambda}_{i}^{\dot{\alpha}}\tilde{\lambda}_{j}^{\dot{\beta}}\epsilon_{\dot{\alpha}\dot{\beta}}$, where $\epsilon$ is the two-index Levi-Civita symbol.

In contrast to the Berkovits-Witten model, the $Z$ and $W$ fields in the 4d ambitwistor model have holomorphic conformal weight $\frac{1}{2}$. Note that the model has a $GL(1)$ symmetry  $\left(Z,W\right)\rightarrow\left(\Omega Z,\Omega^{-1}W\right)$. We will gauge this symmetry as well as the Virasoro symmetry (which contains an $SL(2)$ subgroup). The physical states of the model then correspond to the BRST cohomology. In contrast to ordinary string theory, the spectrum of 4d ambitwistor string only contains field theory degrees of freedom, notably $\mathcal{N}=4$ SYM and CSGR without an infinite tower of massive higher-spin states. In Appendix \ref{brst} we describe the BRST quantization of this model in more detail.

Field theory scattering amplitudes are then obtained from correlation functions of vertex operators corresponding to physical states.  Each vertex operator is described by a pair of complex numbers $\sigma^{\alpha}=\frac{1}{t}(1,\sigma)$, which correspond to homogeneous coordinates on the Riemann sphere at tree-level. For example, an integrated vertex operator encoding the $\mathcal{N}=4$ SYM multiplet (which consists of a gluon, six scalars, and eight fermions) has the following form:
\[
\mathcal{V}_{i}(\sigma)=\int\frac{dt}{t}\delta^{2}\left(\lambda_{i}-t\lambda(\sigma)\right)e^{it\left(\left[\mu(\sigma)\tilde{\lambda}_{i}\right]+\chi(\sigma)\cdot \eta_{i}\right)}j(\sigma)
\]
where $j(\sigma)$ is a Kac-Moody current and $\lambda_i \eta_i$ is the supermomentum. We include subscripts with particle labels to distinguish the external data from worldsheet fields. We also define the vertex operator $\tilde{\mathcal{V}}$ by complex conjugating $\mathcal{V}$ and Fourier transforming back to $\eta$ space. We define the left set $L$ as the set of particles with $\tilde{\mathcal{V}}$ vertex operators, and the right set $R$ as those with $\mathcal{V}$ vertex operators. To compute an $n$-point N$^{k-2}$MHV superamplitude in $\mathcal{N}=4$ SYM is then obtained from the worldsheet correlator 
\[
\left\langle \tilde{\mathcal{V}}_{1}...\tilde{\mathcal{V}}_{k}\mathcal{V}_{k+1}...\mathcal{V}_{n}\right\rangle 
\] 
integrated over the locations of the vertex operators. Particular component amplitudes can then be extracted by integrating out the appropriate $\eta$ variables. At tree-level, we may take all the vertex operators to be integrated and
use the $SL(2)\times GL(1)\sim GL(2)$ symmetry to fix the coordinates
of two vertex operators to be $\sigma^\alpha_{i}=(1,0)$ and $\sigma^\alpha_{j}=(0,1)$. Note that the remaining integral over worldsheet coordinates is localized by the delta functions in the vertex operators which encode the scattering equations. The scattering equations are then refined according to how many particles are in the left and the right set.

As we mentioned above, the 4d ambitwistor string also describes CSGR. The spectrum of this theory contains the following graviton multiplets:
\[
\Phi^{-}=h^{-}\eta_{1}\eta_{2}\eta_{3}\eta_{4}+\eta_{I}\eta_{J}\eta_{K}\psi^{IJK}+\eta_{I}\eta_{J}A^{IJ}+\eta_{I}\psi^{I}+\phi^{-}
\]
\begin{equation}
\Phi^{+}=\phi^{+}\eta_{1}\eta_{2}\eta_{3}\eta_{4}+\eta_{I}\eta_{J}\eta_{K}\psi_{L}\epsilon^{IJKL}+\eta_{I}\eta_{J}A_{KL}\epsilon^{IJKL}+\eta_{I}\psi_{JKL}\epsilon^{IJKL}+h^{+},
\label{gravitonmultiplet}
\end{equation}
where $h^\pm$ refers to helicity $\pm2$, $\left\{ \psi^{IJK},\psi_{IJK}\right\} $ to helicity $\pm3/2$, $\left\{ A^{IJ},A_{IJ}\right\} $ to helicity $\pm1$, $\left\{ \psi^{I},\psi_{I}\right\} $ to helicity $\pm1/2$, and $\phi^\pm$ refer to spin-0 states. Note that the spin-1 states above are distinct from the gluons of $\mathcal{N}=4$ SYM. Also note that the graviton multiplets can have plane wave $e^{i k \cdot x}$ boundary conditions or non-plane wave $A\cdot x e^{i k \cdot x}$ boundary conditions because the equations of motion are fourth order in derivatives. We will present vertex operators corresponding to the graviton multiplet states in the next sections and demonstate their BRST invariance in Appendix \ref{brst}. In contrast to the worldsheet formulae for $\mathcal{N}=4$ SYM and $\mathcal{N}=8$ SUGRA, the scattering equations for CSGR are not in general refined by MHV degree since the left set can contain states from both graviton multiplets. Note that the CSGR spectrum also contains gravitino multiplets consisting of states with helicities $\pm\left\{ \frac{3}{2},1,\frac{1}{2},0,-\frac{1}{2}\right\} $ and plane-wave boundary conditions. The  scattering amplitudes for these states can be computed using the techniques we develop in this paper, although we leave a detailed analysis for future work. For more details about spectrum of CSGR in the context of the Berkovits-Witten model, see \cite{Berkovits:2004jj,Dolan:2008gc}.

\section{Plane Wave Graviton Multiplet Scattering} \label{pw}

In this section, we will consider scattering amplitudes for graviton multiplets with plane wave boundary conditions in CSGR. First we derive a concise worldsheet formula for scalar-graviton amplitudes, and then we lift it to a supersymmetric formula. We denote left set vertex operators with $\tilde{\mathcal{V}}$, and right set with ${\mathcal{V}}$. The scattering equations will then be refined by how many states are in the left set, which will not in general correspond to the MHV degree.

The vertex operators describing gravitons and scalars are given by:
\begin{equation}
\begin{split}
\tilde{\mathcal{V}}^{h^{-}}(\sigma)&=\int\frac{dt}{t^{2}}\left\langle \lambda_{i}\lambda(\sigma)\right\rangle \vertexopleftdelta\\
\tilde{\mathcal{V}}^{\phi^{+}}(\sigma)&=\int t dt \left[\tilde{\lambda}(\sigma)\partial\tilde{\lambda}(\sigma)\right]
\vertexopleftdelta\\
\mathcal{V}^{\phi^{-}}(\sigma)&=\int t dt\left\langle \lambda(\sigma)\partial\lambda(\sigma)\right\rangle
\vertexoprightdelta\\
\mathcal{V}^{h^{+}}(\sigma)&=\int\frac{dt}{t^{2}}\left[\tilde{\lambda}_i\tilde{\lambda}(\sigma)\right]
\vertexoprightdelta.
\end{split}
\label{pwvo}
\end{equation}
We verify the BRST invariance of these vertex operators in appendix \ref{brst}. Let $\Phi^{\pm}$ be the set of positive/negative scalars, and $G^\pm$ be the set of positive/negative helicity gravitons, so that $G^-\cup\Phi^+ = L$ and $G^+\cup\Phi^- = R$. Tree-level graviton-scalar amplitudes can then be obtained from the following correlator
\[
\left\langle \prod_{l_g \in G^- } \tilde{\mathcal{V}}^{h^{-}}_{{l_g}} \prod_{l_\phi \in \Phi^+ } \tilde{\mathcal{V}}^{\phi^{+}}_{l_\phi} \prod_{r_\phi \in \Phi^- } \mathcal{V}^{\phi^{-}}_{r_\phi}\prod_{r_g \in G^+ }\mathcal{V}^{h^{+}}_{r_g} \right\rangle 
\]
integrated over the locations of the vertex operators (modulo $GL(2)$). The correlator can be easily computed using the path integral formalism by combining the exponentials in the vertex operators with the action to obtain the modified Lagrangian
\[
\mathcal{L}=\tilde{\mu}\cdot\bar{\partial}\lambda-\mu\cdot\bar{\partial}\tilde{\lambda}-\tilde{\mu}\cdot\sum_{l\in L}t_{l}\lambda_{l}\delta\left(\sigma-\sigma_{l}\right)+\mu\cdot\sum_{r\in R}t_{r}\tilde{\lambda}_{r}\delta\left(\sigma-\sigma_{r}\right).
\] 
Since the $(\mu, \tilde{\mu})$ fields do not appear anywhere else in the path integral, they can be integrated out yielding delta functionals which localize the functional integrals over the remaining fields onto solutions of the equations of motion
\[
\bar{\partial}\lambda=\sum_{l\in L}t_{l}\lambda_{l}\delta\left(\sigma-\sigma_{l}\right),\,\,\,\bar{\partial}\tilde{\lambda}=\sum_{r\in R}t_{r}\tilde{\lambda}_{r}\delta\left(\sigma-\sigma_{r}\right),
\]
which are uniquely solved by
\begin{equation}
t\lambda(\sigma)=\sum_{l\in L}\frac{\lambda_{l}}{\left(\sigma l\right)},\,\,\, t\tilde{\lambda}(\sigma)=\sum_{r\in R}\frac{\tilde{\lambda}_{r}}{\left(\sigma r\right)},
\label{solutions}
\end{equation}
where $(ij)=\left(\sigma_{i}-\sigma_{j}\right)/t_i t_j$. The amplitude is then given by the following worldsheet integral:
\begin{equation}
\mathcal{A}_{n,|L|}^{h,\phi} =\int\frac{d^{2\times n}\sigma}{GL(2)}\delta(SE)\prod_{l_g \in G^- } H_{l_g} \prod_{l_\phi \in \Phi^+ } \tilde{F}_{l_\phi} \prod_{r_\phi \in \Phi^- } F_{r_\phi}\prod_{r_g \in G^+ }\tilde{H}_{r_g}
\label{sg}
\end{equation}
where $d^{2\times n}\sigma=\Pi_{i=1}^{n}d\sigma_{i}dt_{i}/t_{i}^{3}$,
\begin{equation}
\delta(SE) = \prod_{l\in L}\delta^{2}\left(\tilde{\lambda}_{l}-t_{l}\tilde{\lambda}\left(\sigma_{l}\right)\right) \prod_{r\in R}\delta^{2}\left(\lambda_{r}-t_{r}\lambda\left(\sigma_{r}\right)\right)
\label{scatteqb}
\end{equation}
\[
\tilde{F}_{l}=\sum_{r<r'\in R }\frac{\left[rr'\right]\left(rr'\right)}{\left(lr\right)^{2}\left(lr'\right)^{2}},\,\,\, H_{l}=\sum_{l'\in L,l'\neq l}\frac{\left\langle ll'\right\rangle }{\left(ll'\right)}
\]
\[
F_{r}=\sum_{l<l'\in L}\frac{\left\langle ll'\right\rangle \left(ll'\right)}{\left(rl\right)^{2}\left(rl'\right)^{2}},\,\,\,\tilde{H}_{r}=\sum_{r'\in R,r'\neq r}\frac{\left[rr'\right]}{\left(rr'\right)}.
\]
The delta functions in \eqref{scatteqb} localize the worldsheet integral onto solutions of refined scattering equations.

For $|L|=2$, the scattering equations have only one solution, and it can be found analytically. As we show in Appendix \ref{MHV}, on the support of this solution \eqref{sg} reduces to the Berkovits-Witten result
\begin{equation}
 \mathcal{A}_{n,|L|=2}^{h,\phi} = \delta^{4}(P) \ang{12} ^{4} \prod_{i \in \Phi^+\cup H^+}^{n}\sum_{j=1,j\neq i}^{n}\frac{\left[ij\right]}{\left\langle ij\right\rangle }\frac{\left\langle jx_{i}\right\rangle \left\langle jy_{i}\right\rangle }{\left\langle ix_{i}\right\rangle \left\langle iy_{i}\right\rangle },
\label{bw}
\end{equation}
where $\lambda_{x_{i}}$ and $\lambda_{y_{i}}$
are arbitrary reference spinors. For $|L|>2$ we have verified \eqref{sg} (and its extension to include fermions and spin one states) numerically by matching it against results obtained using Feynman diagrams and color-kinematics duality \cite{Johansson:2017srf} up to eight points with any number of particles in the left set \footnote{We thank Henrik Johansson for providing numerical results derived from color-kinematics duality against which to compare our worldsheet formula.}. In order to do so, new techniques were developed for numerically solving the scattering equations which will be reported on in \cite{Joe}. 

\subsection{Supersymmetric Formula}

The scalar-graviton vertex operators in \eqref{pwvo} can be lifted to the following supersymmetric vertex operators:  
\begin{equation}
\begin{split}
\tilde{\mathcal{V}}^{-}(\sigma)&=\int\frac{dt}{t^{2}}\left\langle \lambda_{i}\lambda(\sigma)\right\rangle \delta^{2|4}\left(\tilde{\lambda}_{i}-t\tilde{\lambda}(\sigma)\right)e^{it\left\langle \tilde{\mu}(\sigma)i\right\rangle }\\
\tilde{\mathcal{V}}^{+}(\sigma)&=\int t dt \left[\tilde{\lambda}(\sigma)\partial\tilde{\lambda}(\sigma)\right] \delta^{2|4}\left(\tilde{\lambda}_{i}-t\tilde{\lambda}(\sigma)\right)e^{it\left\langle \tilde{\mu}(\sigma)i\right\rangle }\\
\mathcal{V}^{-}(\sigma)&=\int t dt\left\langle \lambda(\sigma)\partial\lambda(\sigma)\right\rangle
\delta^{2}\left(\lambda_{i}-t\lambda(\sigma)\right)e^{it\left(\left[\mu(\sigma)i\right]+\chi(\sigma)\cdot\eta_{i}\right)}\\
\mathcal{V}^{+}(\sigma)&=\int\frac{dt}{t^{2}}\left[\tilde{\lambda}_i\tilde{\lambda}(\sigma)\right]\delta^{2}\left(\lambda_{i}-t\lambda(\sigma)\right)e^{it\left(\left[\mu(\sigma)i\right]+\chi(\sigma)\cdot\eta_{i}\right)} 
\end{split}
\label{spwvo}
\end{equation}
where
\[
\delta^{2|4}\left(\tilde{\lambda}_{i}-t\tilde{\lambda}\left(\sigma\right)\right)=\delta^{2}\left(\tilde{\lambda}_{i}-t\tilde{\lambda}\left(\sigma\right)\right)\delta^{4}\left(\eta_{i}-t\tilde{\chi}\left(\sigma\right)\right).
\]
These vertex operators encode all states in the positive/negative helicity graviton multiplets (denoted with a $\pm$), which can occur in the left/right set (denoted with/without a tilde). Hence, CSGR amplitudes are computed using four types of vertex operators, in contrast to $\mathcal{N}=4$ SYM and $\mathcal{N}=8$ SUGRA which have only two types of vertex operators. Note that the $\tilde{\mathcal{V}}^\pm$ vertex operators can be obtained by complex conjugating the $\mathcal{V}^\mp$ vertex operators and Fourier transforming back to $\eta$ space.     

Let us denote the set of states with $\mathcal{V}^\pm$ and $\tilde{\mathcal{V}}^\pm$ vertex operators by $\Phi^\pm$ and $\tilde{\Phi}^\pm$, respectively. Then the left set $L=\tilde{\Phi}^{-}\cup\tilde{\Phi}^{+}$ and the right set $R=\Phi^{-}\cup\Phi^{+}$. A superamplitude is then obtained by computing a correlator of vertex operators integrated over the worldseet (modulo $GL(2)$). As before, one can integrate out the $(\mu, \tilde{\mu})$ fields localizing $(\lambda, \tilde{\lambda})$ onto the solutions in \eqref{solutions}. In the supersymmetric case, we can similarly integrate out the $\chi$ fields, localizing the $\tilde{\chi}$ field onto the following solutions to the equations of motion: 
\[
t\tilde{\chi}(\sigma)=\sum_{r\in R}\frac{\eta_{r}}{\left(\sigma r\right)}.
\]
An $n$-point N$^{k-2}$MHV amplitude with $|L|$ particles in the left set is then given by the following worldsheet formula:
\begin{equation}
\mathcal{A}^n_{k,|L|} =\int\frac{d^{2\times n}\sigma}{GL(2)}\delta(SE)\prod_{l_{-}\in\tilde{\Phi}^{-}}H_{l_{-}}\prod_{l_{+}\in\tilde{\Phi}^{+}}\tilde{F}_{l_{+}}\prod_{r_{-}\in\Phi^{-}}F_{r_{-}}\prod_{r_{+}\in\Phi^{+}}\tilde{H}_{r_{+}},
\label{spw}
\end{equation}
where $k=|\Phi^{-}|+|\tilde{\Phi}^{-}|$ and
\begin{equation}
\delta(SE) = \prod_{l\in L}\delta^{2|4}\left(\tilde{\lambda}_{l}-t_{l}\tilde{\lambda}\left(\sigma_{l}\right)\right) \prod_{r\in R}\delta^{2}\left(\lambda_{r}-t_{r}\lambda\left(\sigma_{r}\right)\right).
\label{scatteq}
\end{equation}
Note that the superamplitude will be unchanged if we replace $\Phi^\pm$ states with $\tilde{\Phi}^\pm$ states as long as $|L|$ is preserved. For the special case where $k=|L|$, a formula in terms of integrals over curves in twistor space was previously conjectured in \cite{Adamo:2012xe}, and it would be interesting to see how this formula is related to \eqref{spw}. Component amplitudes can be extracted by integrating out the appropriate $\eta$ variables. For example, the scalar-graviton amplitudes in \eqref{sg} are obtained by integrating out $\left(\eta_{l}\right)^{4}$ for $l\in L$, and setting $\left(\eta_{r}\right)=0$ for $r \in R$. In a similar way, one can also obtain component amplitudes with fermions and spin-1 states.

\section{Non-Plane Wave Graviton Multiplet Scattering} \label{npwse}

The fourth order equations of motion for conformal gravity lead to a second set of graviton multiplet states with boundary conditions $A\cdot x e^{i k \cdot x}$. Note that if $A \cdot k =0$, then this is actually a solution to the second order wave equation. Following from this we find that vertex operators for non-plane wave states have a free vector index which will be contracted into a choice of $A$ for each state.

We propose the following vertex operators describing non-plane wave gravitons and scalars: 
\begin{equation}
\begin{split}
\tilde{\mathcal{V}}_{h^{-}}^{\alpha \dot{\alpha}}(\sigma)&=\int \frac{dt}{t^2}\left(\lambda_{i}^{\alpha}\mu^{\dot{\alpha}}(\sigma)-\lambda^{\alpha}(\sigma)\frac{\partial}{\partial\lambdat_{i,\dot{\alpha}}}\right)
\vertexopleftdelta\\
\tilde{\mathcal{V}}_{\phi^{+}}^{\alpha \dot{\alpha}}(\sigma)&=\int t dt \left(\partial\tilde{\mu}^{\alpha}(\sigma)\lambdat^{\dot{\alpha}}(\sigma)-\tilde{\mu}^{\alpha}(\sigma)\partial\tilde{\lambda}^{\dot{\alpha}}(\sigma)\right)
\vertexopleftdelta\\
\mathcal{V}_{\phi^{-}}^{\alpha \dot{\alpha}}(\sigma)&=\int t dt \bigg(\lambda^{\alpha}(\sigma)\partial\mu^{\dot{\alpha}}(\sigma)-\partial\lambda^{\alpha}(\sigma)\mu^{\dot{\alpha}}(\sigma)\bigg)
\vertexoprightdelta\\
\mathcal{V}_{h^{+}}^{\alpha \dot{\alpha}}(\sigma)&=\int \frac{dt}{t^2}\left(\tilde{\mu}^{\alpha}(\sigma)\tilde{\lambda}_{i}^{\dot{\alpha}}-\tilde{\lambda}^{\dot{\alpha}}(\sigma)\frac{\partial}{\partial\lambda_{i,\alpha}}\right)
\vertexoprightdelta.
\end{split}
\label{npwvo}
\end{equation}
Since the $(\mu,\tilde{\mu})$ fields appear outside the exponentials, when computing correlation functions we cannot simply combine them with the action and integrate them out as before. On the other hand, if we add source terms for these fields, then we can compute a different correlator where they only appear in the exponentials and obtain the original correlator by taking functional derivatives with respect to the sources and setting them to zero afterwards. In more detail, we add the following source terms to the Lagrangian:  
\[
\mu\cdot \tilde{J}-\tilde{\mu}\cdot J
\]
and consider a correlator of vertex operators like the ones in \eqref{npwvo}, but without $(\mu,\tilde{\mu})$ terms outside the exponenti als. This correlator can then be evaluated by combining the exponentials of the vertex operators with the action and integrating out $(\mu,\tilde{\mu})$ giving rise to delta functionals which localize the $(\lambda, \tilde{\lambda})$ fields onto solutions of the following equations of motion:
\[
\bar{\partial}\lambda=\sum_{l\in L}t_{l}\lambda_{l}\delta\left(\sigma-\sigma_{l}\right)+J,\,\,\,\bar{\partial}\tilde{\lambda}=\sum_{r\in R}t_{r}\tilde{\lambda}_{r}\delta\left(\sigma-\sigma_{r}\right)+\tilde{J},
\]
which are uniquely solved by  
\begin{equation}
t\lambda(\sigma)=\sum_{l\in L}\frac{\lambda_{l}}{(\sigma l)}+\int d\sigma'\frac{tJ\left(\sigma'\right)}{\sigma-\sigma'},\,\,\,t\tilde{\lambda}(\sigma)=\sum_{r\in R}\frac{\tilde{\lambda}_{r}}{(\sigma r)}+\int d\sigma'\frac{tJ\left(\sigma'\right)}{\sigma-\sigma'}.
\label{lambdaJ}
\end{equation}
We can then restore the terms with $(\mu,\tilde{\mu})$ outside the exponentials by taking functional derivatives with respect to $(J,\tilde{J})$ and setting them to zero afterwards (note that after setting the sources to zero, the scattering equations will no longer be deformed). From the explicit form of $(\lambda,\tilde{\lambda})$ in \eqref{lambdaJ}, we conclude that correlators with non-plane wave vertex operators can be evaluated by making the following substitutions for $(\mu,\tilde{\mu})$ outside of the exponentials in \eqref{npwvo}:
\begin{equation}
\mu(\sigma)=\int\frac{d\sigma'}{\sigma-\sigma'}\frac{\delta}{\delta\tilde{\lambda}(\sigma')},\,\,\,\tilde{\mu}(\sigma)=\int\frac{d\sigma'}{\sigma-\sigma'}\frac{\delta}{\delta\lambda(\sigma')}.
\label{replacements}
\end{equation}
These formulae are familiar from canonical quantization.  

Note that the non-plane wave graviton vertex operators in \eqref{npwvo} contain singular terms which cancel out. In the first vertex operator for example, a pole arises in the first term after making the replacement in \eqref{replacements}, but this is precisely cancelled by the pole which arises from $\lambda({\sigma})$ in the second term. Indeed, looking at \eqref{solutions}, we see that the residue of the second pole is $\lambda_i$, which precisely cancels the residue of the first pole. We describe this in more detail in Appendix \ref{npwexamples}, where we also work out some examples at $n$ points with up to two non-plane wave states, and show that amplitudes with non-plane wave states can be obtained by acting on the plane-wave amplitudes with a momentum derivative for each non-plane wave state. This could have been anticipated from the LSZ reduction formula by noting that a non-plane wave solution can be written as a momentum derivative of a plane-wave solution:
\[
A\cdot x\, e^{ik\cdot x}=A\cdot\frac{\partial}{\partial k}\, e^{ik\cdot x},
\]
where $k$ is understood to be off-shell prior to taking the derivative.

Since the amplitudes are manifestly 4d and on-shell, we must define a prescription for taking momentum derivatives of on-shell quantities. We define such a prescription in Appendix \ref{momentumderv}, and use it to derive the following formulae which are sufficient to differentiate any little group invariant function of spinor brackets:
\[
\pd{p^{\dot{\beta}\beta}}\left(\frac{\lambda^{\alpha}}{\ang{\lambda\eta}}\right)=\frac{\eta^{\alpha}}{\ang{\lambda\eta}^{2}}\frac{\lambda_{\beta}\xit_{\dot{\beta}}}{[\xit\lambdat]}
\]
\[
\pdd{(\lambdat^{\dot{\alpha}}\lambda^{\alpha})}{p^{\dot{\beta}\beta}}=\delta_{\beta}^{\alpha}\delta_{\dot{\beta}}^{\dot{\alpha}}-\frac{\xit^{\dot{\alpha}}\xi^{\alpha}\lambda_{\beta}\lambdat_{\dot{\beta}}}{\ang{\lambda\xi}[\xit\lambdat]},
\]
where $\eta$ is an arbitrary spinor and $\xi$ is a reference spinor which parametrizes an off-shell extension of the momentum $k$.
Another subtlety about non-plane wave amplitudes is that they can be expressed in many different ways. For example, using momentum conservation to remove the dependence on the momentum of a particular leg, amplitudes with a single non-plane wave state can be written with a derivative acting only on the momentum-conserving delta function, although the expressions we obtain from worldsheet calculations will generally not be of this form for amplitudes with more than three legs. On the other hand, amplitudes with non-plane wave states are well-defined in a distributional sense. In particular, if we multiply a non-plane wave amplitude by a test function, integrate over momentum space, and perform integration by parts, then we are left with derivatives of the test function times a plane-wave amplitude which is unambiguous. 

Finally, let us point out that non-plane wave scalar-graviton amplitudes can be supersymmetrized using the following vertex operators, which are the non-plane wave analogues of \eqref{spwvo}:
\begin{equation}
\begin{split}
\tilde{\mathcal{V}}_{-}^{\alpha \dot{\alpha}}(\sigma)&=\int \frac{dt}{t^2}\left(\lambda_{i}^{\alpha}\mu^{\dot{\alpha}}(\sigma)-\lambda^{\alpha}(\sigma)\frac{\partial}{\partial\lambdat_{i,\dot{\alpha}}}\right)
 \delta^{2|4}\left(\tilde{\lambda}_{i}-t\tilde{\lambda}(\sigma)\right)e^{it\left\langle \tilde{\mu}(\sigma)i\right\rangle }\\
\tilde{\mathcal{V}}_{+}^{\alpha \dot{\alpha}}(\sigma)&=\int t dt \left(\partial\tilde{\mu}^{\alpha}(\sigma)\lambdat^{\dot{\alpha}}(\sigma)-\tilde{\mu}^{\alpha}(\sigma)\partial\tilde{\lambda}^{\dot{\alpha}}(\sigma)\right)
 \delta^{2|4}\left(\tilde{\lambda}_{i}-t\tilde{\lambda}(\sigma)\right)e^{it\left\langle \tilde{\mu}(\sigma)i\right\rangle }\\
\mathcal{V}_{-}^{\alpha \dot{\alpha}}(\sigma)&=\int t dt \bigg(\lambda^{\alpha}(\sigma)\partial\mu^{\dot{\alpha}}(\sigma)-\partial\lambda^{\alpha}(\sigma)\mu^{\dot{\alpha}}(\sigma)\bigg)
\delta^{2}\left(\lambda_{i}-t\lambda(\sigma)\right)e^{it\left(\left[\mu(\sigma)i\right]+\chi(\sigma)\cdot\eta_{i}\right)}\\
\mathcal{V}_{+}^{\alpha \dot{\alpha}}(\sigma)&=\int \frac{dt}{t^2}\left(\tilde{\mu}^{\alpha}(\sigma)\tilde{\lambda}_{i}^{\dot{\alpha}}-\tilde{\lambda}^{\dot{\alpha}}(\sigma)\frac{\partial}{\partial\lambda_{i,\alpha}}\right)
\delta^{2}\left(\lambda_{i}-t\lambda(\sigma)\right)e^{it\left(\left[\mu(\sigma)i\right]+\chi(\sigma)\cdot\eta_{i}\right)}.
\end{split}
\label{snpwvo}
\end{equation} 
Once again, the superamplitude will depend on both the MHV degree and the number of states in the left set.

\section{Conclusion} \label{conc}
In this paper we investigate tree-level scattering amplitudes of graviton multiplets in CSGR using 4d ambitwistor string theory. This model has the same spectrum as the  Berkovits-Witten twistor string (notably $\mathcal{N}=4$ SYM and a non-minimal version of $\mathcal{N}=4$ CSGR) but gives rise to scattering amplitudes in the form of worldsheet integrals supported on refined scattering equations which are split into two sets, referred to as left and right. In contrast to the 4d ambitwistor string formulae for $\mathcal{N}=4$ SYM and $\mathcal{N}=8$ SUGRA, we find that the scattering equations for CSGR are in general not refined by MHV degree so the amplitudes are labelled by the MHV degree as well as the size of the left set. On the other hand, we are able to obtain very simple formulae for scattering amplitudes which generalize previous results in several ways. 

We obtain a compact formula describing the scattering of any number of scalars and gravitons with any number of particles in the left set. If two particles are in the left set, the worldsheet integrals can be solved analytically reproducing the results of Berkovits and Witten. If more than two particles are in the left set, the worldsheet integrals can be evaluated numerically and we match the results against those obtained using Feynman diagrams and the double copy approach developed in \cite{Johansson:2017srf} up to 8 points with any number of particles in the left set. An explicit algorithm for numerically solving the scattering equations and computing amplitudes with plane wave external states will be described in \cite{Joe}. Moreover we generalize the scalar-graviton amplitudes to a supersymmetric formula using four types of vertex operators which describe states in the left or right set and the positive or negative helicity graviton multiplet.

Since the equations of motion for CSGR are fourth order in derivatives, there are also graviton multiplets with non-plane wave boundary conditions of the form $A\cdot x e^{i k \cdot x}$. Amplitudes with such states are subtle to compute since this requires introducing sources in the worldsheet path integral which lead to deformed scattering equations, as well as developing a prescription for taking momentum derivatives of spinor variables. In the end however, we show that the amplitudes can be obtained by acting on plane wave amplitudes with momentum derivatives.

There are a number of interesting open questions:
\begin{itemize}

\item Conformal symmetry is not manifest in our worldsheet formulae. As explained in \cite{Adamo:2018srx}, this is not surprising since chosing plane wave external states singles out 2-derivative solutions to the 4-derivative equations of motion, breaking conformal invariance. On the other hand, the underlying theory has conformal symmetry so it would be interesting to understand how it is realized at the level of amplitudes. Hidden conformal symmetry of gravitational amplitudes was recently explored in \cite{Loebbert:2018xce}, so it would be interesting to see if the ideas developed in that paper can be applied to conformal gravity.

\item As shown in \cite{Farrow:2017eol}, the 4d ambitwistor string formulae for $\mathcal{N}=4$ SYM and $\mathcal{N}=8$ SUGRA can be mapped into Grassmannian integral formulae which can be derived from a completely different approach involving on-shell diagrams. For $\mathcal{N}=4$ SYM, these formulae suggest a new interpretation of the amplitudes as the volume of a geometric object known as the Amplituhedron \cite{Arkani-Hamed:2013jha}. It would interesting to carry out an analogous mapping for CSGR amplitudes and see if they have a similar geometric interpretation.

\item A double copy construction has recently been proposed for CSGR \cite{Johansson:2017srf}, which involves combining super-Yang-Mills with a certain non-supersymmetric $(DF)^2$ gauge theory, and an ambitwistor string theory describing the latter in general dimensions was proposed in \cite{Azevedo:2017lkz}. It would be interesting to try to formulate the $(DF)^2$ theory using 4d ambitwistor string theory and obtain worldsheet formulae for the scattering amplitudes supported on refined scattering equations.     

\item Classical Einstein gravity in dS$_4$ can be obtained from conformal gravity by imposing Neumann boundary conditions which fix external states to be of the Bunch-Davies form $\left(1+ik\eta\right)e^{-ik\eta+i\vec{k}\cdot\vec{x}}$, where $\eta$ is the conformal time coordinate. These external states have also been used to compute three and four-point de Sitter correlators using Feynman diagrams and BCFW recursion and the results are consistent with holography \cite{Maldacena:2011nz,Raju:2012zs}. Note that the Bunch-Davies state is essentially a linear combination of plane-wave and non-plane wave states which are precisely of the form we have studied in this paper. We therefore hope that the techniques developed in this paper can be used to compute de Sitter correlators using worldsheet methods \footnote{Note that correlators in dS and AdS are related by analytic continuation \cite{Maldacena:2002vr}. A twistor string formula for $\mathcal{N}=8$ SUGRA in AdS$_4$ was proposed in \cite{Adamo:2015ina}, although it is written in terms of external states which make it difficult to relate it to results obtained using other methods.}. 

\end{itemize}

In summary, we find that 4d ambitwistor string theory reveals interesting new mathematical structure in the scattering amplitudes of CSGR, which appears to be very different from the structure previously found in $\mathcal{N}=4$ SYM and $\mathcal{N}=8$ SUGRA. We hope that exploring the directions described above will lead to a deeper understanding of gravitational amplitudes which can ultimately be applied to more realistic models.

\begin{center}
\textbf{Acknowledgements}
\end{center}

We thank Simon Badger, Nathan Berkovits, and Paul Heslop for useful discussions, and especially Tim Adamo for providing comments on the manuscript and Henrik Johansson for sharing numerical results. AL is supported by the Royal Society as a Royal Society University Research Fellowship holder, and JF is funded by EPSRC PhD scholarship EP/L504762/1. 

\appendix

\section{BRST Quantization} \label{brst}
In this appendix, we will review how to BRST quantize the 4d ambitwistor
string theory in \eqref{action}. After gauging the $GL(1)$ symmetry $\left(Z,W\right)\rightarrow\left(\Omega Z,\Omega^{-1}W\right)$
and introducing a worldsheet vielbein by taking $\bar{\partial}\rightarrow\bar{\partial}+e\partial$, the Lagrangian becomes 
\[
\mathcal{L}\rightarrow\mathcal{L}+aZ\cdot W+eT_{matter},
\]
where $a$ is a worldsheet gauge field and

\[
T_{matter}=\frac{1}{2}\left(Z \cdot \partial W-W \cdot \partial Z\right)+T_{j},
\]
where $T_{j}$ is the current algebra stress tensor. 

Note that this action is a $\left(\beta,\gamma\right)$ ghost system with holomorphic conformal weights $(1/2,1/2)$. A general $\left(\beta,\gamma\right)$ system with holomorphic conformal
weights $(\lambda,1-\lambda)$ has the stress tensor 

\[
T_{\beta\gamma}=\lambda\beta\partial\gamma-\epsilon(1-\lambda)\gamma\partial\beta,
\]
where $\epsilon=\pm1$ for bosonic/fermionic statistics. The central charge can then
be read off from the OPE of $T$ with itself and is given by
\begin{equation}
c=2\epsilon\left(6\lambda^{2}-6\lambda+1\right).
\label{central}
\end{equation}

We may gauge-fix $e=a=0$ using the Fadeev-Popov procedure by introducing
ghost systems $(b,c)$ and $\left(\tilde{b},\tilde{c}\right)$ with holomorphic
conformal weights $(2,-1)$ and $(1,0)$, respectively. The stress tensor for
the ghosts is then given by $T_{ghost}=T_{bc}+T_{\tilde{b}\tilde{c}}$
where 
\[
T_{bc}=2b\partial c-c\partial b,\,\,\, T_{\tilde{b}\tilde{c}}=\tilde{b}\partial\tilde{c}.
\]
Using \eqref{central}, the contribution of the ghosts to the central charge is $c_{ghost}=-26-2=-28$.
We then define the BRST charge $Q$ as follows:
\[
Q=\oint d\sigma\left(c\left(T_{matter}+T_{ghost}\right)+\tilde{c}Z\cdot W\right).
\]
The key property that $Q$ must satisfy is nilpotency, i.e. $Q^{2}=0$.
In order for $Q$ to satisfy this constraint, the total central charge
must vanish. The $(Z,W)$ system has zero central charge since the bosonic
contributions cancel the fermionic ones, so the central charge of the
current algebra must be $c_{j}=+28$. 

The physical states of the theory correspond to the BRST cohomology. Hence, the corresponding vertex operators must be $Q$-closed, i.e. $\left\{ Q,\mathcal{V}\right\} =0$. The condition of $Q$-closure implies
that the vertex operators must have holomorphic conformal weight $w_{\mathcal{V}}=1$
and $GL(1)$ weight $q_{\mathcal{V}}=0$. The conformal and $GL(1)$
weights may in turn be read off from the OPE of the vertex operator with $T$ and $Z\cdot W$:
\[
T(\sigma)\mathcal{V}(\sigma')=\frac{w_{\mathcal{V}}\mathcal{V}(\sigma)}{\left(\sigma-\sigma'\right)^{2}}+...,\,\,\, Z\cdot W(\sigma)\mathcal{V}(\sigma')=\frac{q_{\mathcal{V}}\mathcal{V}(\sigma)}{\sigma-\sigma'}+...
\]
where the ellipsis denote less singular terms. 

Let us verify that the vertex operators considered in this paper are
$Q$-closed. Since the equations of motion for conformal gravity are
fourth order in derivatives, the spectrum contains plane wave states
of the form $e^{ik\cdot x}$ as well as non-plane wave states of the
form $A\cdot xe^{ik\cdot x}$. Moreover, in the ambitwistor string
framework, vertex operators with opposite helicity are simply complex
conjugates of each other and are therefore naturally divided into two
sets, which we shall refer to as left and right sets (the scattering
equations are then refined by the number of states in each set). 

A plane-wave vertex operator in the right set is schematically of
the form
\[
\delta^{2}\left(\lambda_{i}-t\lambda(\sigma)\right)e^{it\mu(\sigma)\cdot\tilde{\lambda}_{i}},
\]
where $k_i=\lambda_{i}\tilde{\lambda}_{i}$ is the on-shell momentum.  
Using the incidence relation adapted to worldsheet fields $\mu(\sigma)=x\cdot\lambda(\sigma)$
(where $x$ is a point in spacetime), we see that the exponential
reduces to a plane-wave on the support of the delta function. Let us therefore consider
an ansatz for a plane-wave vertex operator of the form
\begin{equation}
\mathcal{V}(\sigma)=\int\frac{dt}{t^{\gamma}}\delta^{2}\left(\lambda_{i}-t\lambda(\sigma)\right)e^{it\mu(\sigma)\cdot\tilde{\lambda}_{i}}\left[\tilde{\lambda}(\sigma)\tilde{\lambda}_{i}\right]^{s-1}j(\sigma)\label{eq:s=00003D1,2},
\end{equation}
where $s\geq1$. In practice, one can avoid a tedious (but straightforward)
OPE calculation using the following rules for computing conformal
and $GL(1)$ weights: 
\[
T:\,\,\,[Z]=[W]=-[t]=\frac{1}{2},\,\,\,[\partial]=1
\]
\[
Z\cdot W:\,\,\,[Z]=-[W]=-[t]=1,\,\,\,[\partial]=0,
\]
where weights of $t$ are fixed by the consistency condition that
$[t Z]=0$ (for vertex operators in the left set, $t$ will have opposite
weights). We take external
spinors have zero weight i.e. $\left[\lambda_{i}\right]=\left[\tilde{\lambda}_{i}\right]=0$.
Applying these rules to the vertex operator in \eqref{eq:s=00003D1,2},
we find that

\[
w_{\mathcal{V}}=\frac{1}{2}\left(\gamma-1\right)+\frac{1}{2}(s-1)+w_{j},\,\,\, q_{\mathcal{V}}=(\gamma-1)-(s-1),
\]
where $w_{j}$ is the conformal weight of the current algebra.
$Q$-closure then implies that $\gamma=s$ and $w_{j}=2-s$, which
implies that $s\leq2$ \footnote{Note that if we do not impose the constraint $q_\mathcal{V}=0$, i.e. we do not gauge the $GL(1)$ symmetry of the worldsheet theory, then vertex operators with higher spin appear to be allowed.}. If $s=1$, then the vertex operator reduces to 
\begin{equation}
\mathcal{V}(\sigma)=\int\frac{dt}{t}\delta^{2}\left(\lambda_{i}-t\lambda(\sigma)\right)e^{it\mu(\sigma)\cdot\tilde{\lambda}_{i}}j(\sigma)\label{eq:s=00003D1}
\end{equation}
which describes a gluon in $\mathcal{N}=4$ SYM. For $s=2$, the vertex
operator describes a graviton with plane wave boundary conditions:
\[
\mathcal{V}(\sigma)=\int\frac{dt}{t^{2}}\delta^{2}\left(\lambda_{i}-t\lambda(\sigma)\right)e^{it\mu(\sigma)\cdot\tilde{\lambda}_{i}}\left[\tilde{\lambda}(\sigma)\tilde{\lambda}_{i}\right],
\]
where we discarded the current algebra since $w_j=0$.

To deduce the vertex operator for a scalar in the left set, consider
the ansatz
\begin{equation}
\mathcal{V}(\sigma)=\int\frac{dt}{t^{\gamma}}\delta^{2}\left(\lambda_{i}-t\lambda(\sigma)\right)e^{it\mu(\sigma)\cdot\tilde{\lambda}_{i}}\left\langle \lambda(\sigma)\partial\lambda(\sigma)\right\rangle ^{1-s}j(\sigma)\label{eq:s=00003D0,1}
\end{equation}
where $s\leq1$. Using the rules described above, one finds that
\[
w_{\mathcal{V}}=\frac{1}{2}\left(\gamma-1\right)+2(1-s)+w_{j},\,\,\, q_{\mathcal{V}}=2(1-s)+(\gamma-1).
\]
Imposing $w_{\mathcal{V}}=1$ and $q_{\mathcal{V}}=0$ then implies
that $\gamma=2s-1$ and $w_{j}=s$, from which we deduce that $s\geq0$.
If $s=1$, then the vertex operator reduces to the gluon vertex operator
in \eqref{eq:s=00003D1}, but if $s=0$ it describes a scalar with plane-wave boundary conditions
\[
\mathcal{V}(\sigma)=\int tdt\delta^{2}\left(\lambda_{i}-t\lambda(\sigma)\right)e^{it\mu(\sigma)\cdot\tilde{\lambda}_{i}}\left\langle \lambda(\sigma)\cdot\partial\lambda(\sigma)\right\rangle .
\]

Let us now turn our attention to non-plane wave states. Let us consider
the following ansatz:

\[
\mathcal{V}(\sigma)=\int\frac{dt}{t^{\gamma}}\left(A_{\alpha\dot{\alpha}}\left(\tilde{\mu}^{\alpha}(\sigma)\tilde{\lambda}_{i}^{\dot{\alpha}}-\tilde{\lambda}^{\dot{\alpha}}(\sigma)\frac{\partial}{\partial\lambda_{i,\alpha}}\right)\right)^{s-1}\delta^{2}\left(\lambda_{i}-t\lambda(\sigma)\right)e^{it\mu(\sigma)\cdot\tilde{\lambda}_{i}}.
\]
Following an analysis very similar to the one for \eqref{eq:s=00003D1,2},
we find that $s=\gamma=2$, so the vertex operator reduces to
that of a non-plane wave graviton. Similarly, we find that the following ansatz 
\[
\mathcal{V}(\sigma)=\int\frac{dt}{t^{\gamma}}\left(A_{\alpha\dot{\alpha}}\left(\lambda^{\alpha}(\sigma)\partial\mu^{\dot{\alpha}}(\sigma)-\mu^{\dot{\alpha}}(\sigma)\partial\lambda^{\alpha}(\sigma)\right)\right)^{1-s}\delta^{2}\left(\lambda_{i}-t\lambda(\sigma)\right)e^{it\mu(\sigma)\cdot\tilde{\lambda}_{i}}
\]
must satisfy $s=0$ and $\gamma=-1$, and reduces to the vertex operator
for a non-plane wave scalar.

\section{Derivation of Berkovits-Witten Formula} \label{MHV}

In this appendix, we will evaluate the worldsheet integral in \eqref{sg} for the case $|L|=2$. In this case, the worldsheet integral is straightforward to evaluate analytically. Let us take the left set to be $L=\left\{ 1,2\right\}$ and the right set to be $R=\left\{ 3,...,n\right\}$. Using the $GL(2)$ symmetry to set $\sigma_1^\alpha=(1,0)$ and $\sigma_2^\alpha=(0,1)$, the delta functions encoding the refined scattering equations reduce to
\[
\prod_{l\in L}\delta^{2}\left(\tilde{\lambda}_{l}-t_{l}\tilde{\lambda}\left(\sigma_{l}\right)\right)=\left\langle 12\right\rangle ^{2}\delta^{4}(P),
\]
\[
\delta^{2}\left(\lambda_{r}-t_{r}\lambda\left(\sigma_{r}\right)\right)=\frac{\left(1r\right)^{2}\left(2r\right)^{2}}{\left\langle 12\right\rangle (12)}\delta\left(\sigma_{r}^{1}-\frac{\left\langle 21\right\rangle }{\left\langle r1\right\rangle }\right)\delta\left(\sigma_{r}^{2}-\frac{\left\langle 21\right\rangle }{\left\langle r2\right\rangle }\right),\,\,\, r\in R.
\]
The remaining worldsheet integrals then localize onto the solution
\begin{equation}
\sigma_{r}^{\alpha}=\left(\frac{\left\langle 21\right\rangle }{\left\langle r1\right\rangle },\frac{\left\langle 21\right\rangle }{\left\langle r2\right\rangle }\right),\,\,\, r\in R.
\label{msol}
\end{equation}
On the support of the refined scattering equations, \eqref{sg} then reduces to
\begin{equation}
\begin{split}
\mathcal{A}_{n,|L|=2}^{h,\phi} &= \delta^4(P)\ang{12}^2 \prod_{r \in R} \frac{(1r)^2(2r)^2}{\ang{12}(12)}
 \prod_{l_g\in G^-}\frac{\ang{12}}{(12)}\prod_{l_\phi\in \Phi^+}\sum_{r<r' \in R}\frac{[rr'](rr')}{(l_\phi r)^2(l_\phi r')^2}
 \\&\hspace{3cm}
 \prod_{r_\phi\in \Phi^-}\frac{\ang{12}(12)}{(1r_\phi)^2(2r_\phi)^2}\prod_{r_g\in G^+}\sum_{r\in R}\frac{[r_g r]}{(r_g r)}\\
 &= \delta^4(P)\frac{\ang{12}^4}{(12)^2} \prod_{l_\phi\in \Phi^+}\sum_{r<r' \in R}\frac{[rr'](rr')(12)}{(l_\phi r)^2(l_\phi r')^2\ang{12}} 
 \prod_{r_g\in G^+}\sum_{r\in R}\frac{[r_g r](1r_g)^2(2r_g)^2}{(r_g r)\ang{12}(12)}.
\end{split}
\end{equation}
Plugging in \eqref{msol}, we find that the factor associated with each $h^+$ leg is given by
\begin{equation}
\sum_{r\in R}\frac{[r_g r](1r_g)^2(2r_g)^2}{(r_g r)\ang{12}(12)} = \sum_{r\in R}\frac{[r_g r]\ang{1r}\ang{2r}}{\ang{r_g r}\ang{1r_g}\ang{2r_g}} = \psi_{r_g,n}^{\lambda_1\lambda_2},
\end{equation}
where we define the gravitational inverse soft factor for leg $j$ as follows:
\begin{equation}
\psi_{j,n}^{ab}=\sum_{k=1,k\neq j}^{n}\frac{\left[jk\right]}{\left\langle jk\right\rangle }\frac{\left\langle ka\right\rangle \left\langle kb\right\rangle }{\left\langle ja\right\rangle \left\langle jb\right\rangle },
\label{gisf}
\end{equation}
where $a$ and $b$ are reference spinors. Using momentum conservation, it is possible to show that the inverse soft factor is independent of the choice of reference spinors so we will just refer to it as $\psi_{j,n}$.
Similarly, we find that the factor associated with each $\phi^+$ leg is given by
\begin{equation}
\sum_{r<r' \in R}\frac{[rr'](rr')(12)}{(l_\phi r)^2(l_\phi r')^2\ang{12}} =\psi_{l_\phi,n}, 
\end{equation}
where we checked the second equality numerically up to high multiplicity. Hence, we find that for $|L|=2$, \eqref{sg} reduces to the formula of Berkovits and Witten;
\[
\mathcal{A}_{n,|L|=2}^{h,\phi}  = \delta^4(P)\ang{12}^4 \prod_{i\in \Phi^+\cup G^+}\psi_{i,n}.
\]

\section{Momentum Derivatives} \label{momentumderv}

We would like to be able to take derivatives of on-shell quantities with respect to off-shell momenta. As written, this problem is not well specified as there are four degrees of freedom in an off shell momentum $p$ but only three degrees of freedom in the spinor variables which parametrize an on-shell momentum after removing the little group redundancy. We work with real momenta such that $\lambdat^{\dot{\beta}} = \lambda^{\beta*}$. To make the problem well-defined, we will introduce a fourth degree of freedom $\alpha$ in addition to the three degrees of freedom in the spinor variables and write the off-shell momentum in terms of $(\lambda^\beta, \alpha)$ as follows:
\begin{equation}
\label{eqn:coordtrans}
p^{\dot{\beta}\beta} = \lambdat^{\dot{\beta}}\lambda^\beta + \alpha \tilde{\xi}^{\dot{\beta}}\xi^\beta
\end{equation}
where $\xi^{\alpha\dot{\alpha}}=\xi^{\alpha}\tilde{\xi}^{\dot{\alpha}}$ is a reference null vector which encodes an off-shell direction.

Inverting these equations to solve for $\alpha(p)$ is simple, and we find that $\alpha(p) = \frac{p^2}{2 p \cdot \xi}$. We can also solve for $\lambda(p)^\beta$. To see this, contract with the spinor $\xit^{\dot{\beta}}$ to arrive at 
$\xit_{\dot{\beta}} p^{\dot{\beta}\beta} = [\xit\lambdat]\lambda^\beta$. For real momenta it is clear that $\left|[\xit\lambdat]\right|^2 = \xi \cdot p$, hence there must exist some phase $\theta(p)$ such that $[\xit\lambdat] = e^{-i \theta} \sqrt{\xi\cdot p}$. Given that $\lambda$ is defined only up to an arbitrary phase, we can express it in terms of $p$ as follows:

\begin{equation}
\lambda(p)^\beta =e^{i \theta(p)} \frac{\xit_{\dot{\beta}}p^{\dot{\beta}\beta}}{\sqrt{\xi\cdot p}}.
\label{dl}
\end{equation}
Note that for the choice $\xi^\alpha = \begin{pmatrix}0\\1\end{pmatrix}$ and $\theta=0$ we recover the well-known expression

\begin{equation}
\lambda(p)^\beta =  \frac{1}{\sqrt{p^0+p^3}}\begin{pmatrix}p^0+p^3\\p^1-ip^2\end{pmatrix}. 
\end{equation}
Differentiating \eqref{dl} with respect to the off-shell momentum then gives
\begin{equation}
\begin{split}
\label{eqn:dlamdp}
\pdd{\lambda(p)^\alpha}{p^{\dot{\beta}\beta}} &= \pd{p^{\dot{\beta}\beta}}\left(\frac{e^{i\theta(p)}\xit_{\dot{\alpha}}p^{\dot{\alpha}\alpha}}{\sqrt{\xi\cdot p}}\right) \\
&= \frac{\delta^\alpha_\beta e^{i\theta(p)}\xit_{\dot{\beta}}}{\sqrt{\xi\cdot p}} - \frac{1}{2}\frac{e^{i\theta(p)}\xit_{\dot{\alpha}}p^{\dot{\alpha}\alpha}}{\sqrt{\xi\cdot p}}\frac{\xi_{\beta}\xit_{\dot{\beta}}}{\xi\cdot p} + i \pdd{\theta(p)}{p^{\dot{\beta}\beta}}\frac{e^{i\theta(p)}\xit_{\dot{\alpha}}p^{\dot{\alpha}\alpha}}{\sqrt{\xi\cdot p}} \\
&= \frac{\delta^\alpha_\beta \xit_{\dot{\beta}}}{[\xit\lambdat]} - \frac{1}{2}\frac{\lambda(p)^\alpha\xi_{\beta}\xit_{\dot{\beta}}}{\ang{\lambda\xi}[\xit\lambdat]} + i \pdd{\theta(p)}{p^{\dot{\beta}\beta}}\lambda(p)^\alpha.
\end{split}
\end{equation}

In general, differentiating a function of spinor brackets which is not little-group invariant will result in a $\pdd{\theta(p)}{p}$ term. Let us therefore consider momentum derivatives of the little-group invariants $\lambdat^{\dot{\alpha}}\lambda^\alpha$ and $\frac{\lambda^\beta}{\ang{\lambda \eta}}$, where $\eta$ is an arbitrary spinor (which can come from a different external leg for example). We find that
\begin{equation}
\begin{split}
\label{eqn:ddplamonang}
\pd{p^{\dot{\beta}\beta}} \left(\frac{\lambda^\alpha}{\ang{\lambda \eta}}\right) &= \frac{\ang{\lambda \eta} \pdd{\lambda^\alpha}{p^{\dot{\beta}\beta}} + \eta_\gamma \pdd{\lambda^\gamma}{p^{\dot{\beta}\beta}} \lambda^\alpha}{\ang{\lambda \eta}^2} \\
&= \frac{\eta^\alpha}{\ang{\lambda \eta}^2}\lambda_\gamma\pdd{\lambda^\gamma}{p^{\dot{\beta}\beta}} = \frac{\eta^\alpha}{\ang{\lambda \eta}^2}\frac{\lambda_\beta\xit_{\dot{\beta}}}{[\xit\lambdat]},
\end{split}
\end{equation}
where in the second line we used the Schouten identity, and in the third equality we used \eqref{eqn:dlamdp}. Furthermore, using equation~(\ref{eqn:coordtrans}) we find that

\begin{equation}
\begin{split}
\pdd{(\lambdat^{\dot{\alpha}}\lambda^\alpha)}{p^{\dot{\beta}\beta}} &= \pd{p^{\dot{\beta}\beta}}\left(p^{\dot{\alpha}\alpha} -  \frac{p^2}{2\xi\cdot p} \xi^{\dot{\beta}}\xi^\beta \right).
\end{split}
\end{equation}
This calculation involves only derivatives of vectors with respect to vectors, and hence we do not need equation~(\ref{eqn:dlamdp}). The result is

\begin{equation}
\begin{split}
\label{eqn:ddplamtlam}
\pdd{(\lambdat^{\dot{\alpha}}\lambda^{\alpha})}{p^{\dot{\beta}\beta}}=\delta_{\beta}^{\alpha}\delta_{\dot{\beta}}^{\dot{\alpha}}-\frac{\xit^{\dot{\alpha}}\xi^{\alpha}\lambda_{\beta}\lambdat_{\dot{\beta}}}{\ang{\lambda\xi}[\xit\lambdat]}.
\end{split}
\end{equation}
Note that the right-hand-side is a projection matrix which removes components along the off-shell direction, i.e. $\xi^{\dot{\beta}}\xi^\beta\pdd{\lambdat^{\dot{\alpha}}\lambda^\alpha}{p^{\dot{\beta}\beta}} = 0$ and $p^{\dot{\beta}\beta}\pdd{\lambdat^{\dot{\alpha}}\lambda^\alpha}{p^{\dot{\beta}\beta}} = \lambdat^{\dot{\alpha}}\lambda^\alpha$.  

\section{Non-plane Wave Examples} \label{npwexamples}

In this section, we will work out examples of scattering amplitudes for non-plane wave states of the form $A\cdot x e^{ik\cdot x}$ using the vertex operators proposed in section \ref{npwse}, and use the method described in Appendix \ref{momentumderv} to express them as momentum derivatives of plane wave amplitudes.

\subsection{Non-plane Wave Scalar}

We will first calculate an amplitude with two plane wave negative helicity gravitons, a negative multiplet scalar with non-plane wave boundary conditions, and $n-3$ plane wave positive helicity gravitons. We then define the left set to be $L = \{1,2\}$, the right set $R$ to contain the remaining particles, and the set $R' = \{4, ... n\}$ to be the set of positive helicity gravitons. The vertex operators for these states can be found in \eqref{pwvo} and \eqref{npwvo}.

After replacing $(\mu,\tilde{\mu})$ outside the exponential in the non-plane wave scalar vertex operator and taking functional derivatives according to \eqref{replacements}, we obtain the following worldsheet formula:
\begin{equation}
\begin{split}
\mathcal{A}(h^-h^-\phi^-_x h^+ ... h^+) &= 
A_{3\,\alpha\dot{\alpha}}\int\sigmeasure\frac{\ang{12}^{2}}{(12)^{2}}\Bigg(\Big(\prod_{\rho\in R'}\sum_{r\in R}\frac{\squ{\rho r}}{(\rho r)}\Big)\frac{(12)}{(13)^{2}(23)^{2}}\left(\bra{2}^{\alpha}\frac{\partial}{\partial\lambdat_{1\dot{\alpha}}}-\bra{1}^{\alpha}\frac{\partial}{\partial\lambdat_{2\dot{\alpha}}}\right)\\
&\hspace{2cm}+  \sum_{\rho \in R',l \in L}\frac{\sket{\rho}^{\dot{\alpha}}\bra{l}^{\alpha}(\rho l)}{(3 \rho)^2 (3 l)^2}\Big( \prod_{\rho' \in R', \rho' \neq \rho} \sum_{r\in R}\frac{\squ{\rho' r}}{(\rho' r)} \Big)
 \Bigg) \delta(SE).
 \end{split}
\end{equation}
The first term comes from acting with the functional derivatives on the delta functions imposing the scattering equations, and the second term comes from acting on the spinor brackets $\left[\tilde{\lambda}_{\rho}\tilde{\lambda}\left(\sigma_{\rho}\right)\right]$ in the positive-helicity graviton vertex operators. 

We can evaluate the worldsheet integral analytically following the same procedure described in Appendix \ref{MHV}. Using the $GL(2)$ symmetry to fix $\sigma_1^\alpha=(1,0)$ and $\sigma_2^\alpha=(0,1)$ and converting the delta functions in the left set into a momentum conserving delta function, we see that the remaining terms do not depend on $\sket{1}$ or $\sket{2}$. Furthermore, the Jacobian from the scattering equation delta functions only contains angle brackets, so the $\frac{\partial}{\partial \lambdat}$ will act only on the momentum conserving delta function. We can then simplify this part of the calculation as follows:

\begin{equation}
\begin{split}
\left(\frac{\partial}{\partial \lambdat_{1\dot{\alpha}}}\bra{2}^{\alpha} - \frac{\partial}{\partial \lambdat_{2\dot{\alpha}}}\bra{1}^{\alpha} \right)\delta^4(P) 
&= \left(\bra{1\beta}\pd{P_{\dot{\alpha}\beta}}\bra{2}^{\alpha} - \bra{2\beta}\pd{P_{\dot{\alpha}\beta}}\bra{1}^{\alpha}\right)\delta^4(P) \\
&= \ang{12}\pd{P_{3\,\dot{\alpha}\alpha}}\delta^4(P).
 \end{split}
\end{equation}
After some further simplification using the Schouten identity, we obtain 

\begin{equation}
\begin{split}
\mathcal{A}(h^{-}h^{-}\phi_{x}^{-}h^{+}...h^{+})&=\ang{12}^{4}A_{3\,\alpha\dot{\alpha}}\Bigg(\prod_{\rho\in R'}\psi_{\rho,n}^{\ket{1}\ket{2}}\pd{P_{3\,\dot{\alpha}\alpha}}
\\
&\hspace{2cm} +\sum_{\rho\in R'} 
\frac{\sket{\rho}^{\dot{\alpha}}\left(\ang{12}\ang{\rho 3}\bra{3}^{{\alpha}}+\ang{23}\ang{13}\bra{\rho}^{{\alpha}}\right)}{\ang{3\rho}^2\ang{1\rho}\ang{2\rho}}
\prod_{\rho' \in R', \rho' \ne \rho}\psi_{\rho',n} \Bigg)\delta^4(P),
\end{split}
\label{hhphi}
\end{equation}
where the gravitational inverse soft factor $\psi_{\rho,n}$ was defined in \eqref{gisf}.

Using the results in Appendix \ref{momentumderv}, it is not difficult to show that \eqref{hhphi} can be obtained by applying a momentum derivative to a plane-wave amplitude as follows:
\begin{equation}
\begin{split}
\mathcal{A}(h^{-}h^{-}\phi_{x}^{-}h^{+}...h^{+})=\ang{12}^{4}A_3 \cdot \pd{P_3}\left(\prod_{\rho\in R'}\psi_{\rho,n}\delta^{4}(P)\right).
\end{split}
\label{smderv}
\end{equation}
Clearly for $n=3$, $|R'| = 0$ and the result holds. To show this for $n>3$, let us compute the momentum derivative of the gravitational inverse-soft factor for leg $j$  with respect to particle $i$ where $i \neq j$ and assume that the reference spinors do not depend on $i$. We then find that 
\begin{equation}
\begin{split}
\label{eqn:dervisf1}
\pd{P_i^{\dot{\beta}\beta}} \psi_{j,n}^{ab} &= \pd{P_i^{\dot{\beta}\beta}}\sum_{k \neq j}\frac{\squ{jk}}{\ang{jk}}\frac{\ang{ka}\ang{kb}}{\ang{ja}\ang{jb}}\\
&= \frac{1}{\ang{ja}\ang{jb}} \left( \lambdat_{j{\dot{\alpha}}}\pd{P_i^{\dot{\beta}\beta}}\left(\lambdat_i^{\dot{\alpha}}\lambda_i^{\alpha}\right)a_{\alpha} \frac{\ang{ib}}{\ang{ji}}   
		+ \squ{ji}\ang{ia} \pd{P_i^{\dot{\beta}\beta}}\left(\frac{\lambda_i^{\alpha}}{\ang{ji}}\right)b_{\alpha}\right)\\
&= \frac{\lambdat_{j\dot{\beta}}}{\ang{ja}\ang{jb}\ang{ij}^2} \bigg(\ang{ab}\ang{ji}\lambda_{i\beta} + \ang{ia}\ang{bi}\lambda_{j\beta}\bigg),
\end{split}
\end{equation}
where we used equations~(\ref{eqn:ddplamonang}) and~(\ref{eqn:ddplamtlam}) and chose the reference spinor to be $\xit^{\dot{\alpha}} = \lambdat_j^{\dot{\alpha}}$. Setting $i=\rho$, $j=3$, $a=\lambda_1$, and $b=\lambda_2$, we see that the second term in \eqref{hhphi} contains the derivative of $\psi_{\rho,n}$, from which \eqref{smderv} follows.

\subsection{Non-plane Wave Graviton} 

We now compute an amplitude with one negative-helicity non-plane wave graviton, $\mathcal{A}(h^-_x h^- h^+ ... h^+)$. As outlined in section \ref{npwse}, the vertex operator for $h^-_x$ has divergences which cancel, and we show the details of this here. Following the same steps as in the previous section, we obtain the following worldsheet formula: 
\begin{equation}
\begin{split}
\mathcal{A}(h^-_x h^- h^+ ... h^+) &= 
 A_{1 \,\beta \dot{\beta}}\int \sigmeasure \frac{\ang{12}}{(12)} \\
 &
\Bigg( \lim_{\alpha\rightarrow\sigma_1}
 \Bigg(
 \frac{\ket{1}^\beta\pd{\lambdat_{1{\dot{\beta}}}}}{\alpha-\sigma_1} + \frac{\ket{1}^\beta\pd{\lambdat_{2{\dot{\beta}}}}}{(12)} - \frac{\ket{1}^\beta\pd{\lambdat_{1{\dot{\beta}}}}}{\alpha-\sigma_1} - \frac{\ket{2}^\beta\pd{\lambdat_{1{\dot{\beta}}}}}{(12)}
 \Bigg)\prod_{r \in R} \sum_{r'\in R}\frac{\squ{r r'}}{(r r')} \\
 & \hspace{4cm}+\sum_{r \in R} \frac{\ket{1}^{{\beta}}\sbra{r}^{\dot{\beta}}}{(1r)}\prod_{r' \neq r \in R} \sum_{r''\in R}\frac{\squ{r' r''}}{(r' r'')} 
  \Bigg) 
 \delta(SE)
 \end{split}
\end{equation}
where the first term comes from acting with the functional and ordinary derivative in $h^-_x$ on the delta functions imposing the scattering equations, the second term comes from acting with the functional derivative on the spinor brackets $\left[\tilde{\lambda}_{r}\tilde{\lambda}\left(\sigma_{r}\right)\right]$ in the positive-helicity graviton vertex operators, and we have regulated divergent terms by taking a limit.  

Cancelling the singular terms and carrying out the worldsheet integral as described in Appendix \ref{MHV} then gives
\begin{equation}
\begin{split}
\mathcal{A}(h^-_x h^- h^+ ... h^+) &= 
A_{1\, \beta \dot{\beta}} \int \sigmeasure \frac{\ang{12}}{(12)} 
\Bigg( \Bigg(
 \frac{\ket{1}^\beta\pd{\lambdat_{2{\dot{\beta}}}}- \ket{2}^\beta\pd{\lambdat_{1{\dot{\beta}}}}}{(12)}
 \Bigg)\prod_{r \in R} \sum_{r'\in R}\frac{\squ{r r'}}{(r r')} \\
& \hspace{2cm}+\sum_{r \in R} \frac{\ket{1}^\beta\sbra{r}^{\dot{\beta}}}{(1r)}\prod_{r' \neq r \in R} \sum_{r''\in R}\frac{\squ{r' r''}}{(r' r'')} 
  \Bigg) 
 \delta(SE)\\
&= \ang{12}^4 A_{1\, \beta \dot{\beta}}\Bigg( \prod_{r \in R}\psi_{r,n}^{\ket{1}\ket{2}} \pd{P_{1\,\beta{\dot{\beta}}}}  +\sum_{r\in R} 
\frac{\ang{12}\ket{1}^\beta\sbra{r}^{\dot{\beta}}}{\ang{1r}^2\ang{2r}}
\prod_{r' \in R, r' \ne r}\psi_{r',n} \Bigg)\delta^4(P).
 \end{split}
\end{equation}
Using the results from Appendix~\ref{momentumderv} to differentiate the gravitational inverse soft factor as we did in the previous section, we arrive at the final result that
\begin{equation}
\begin{split}
 \mathcal{A}(h^-_x h^- h^+ ... h^+)  = \ang{12}^4 A_1\!\cdot \!\pd{P_{1}} \Bigg(\prod_{r \in R}\psi_{r,n}^{\ket{1}\ket{2}} \delta^4(P)\Bigg).
 \end{split}
\end{equation}
Following a similar calculation with two $h^-_x$ states we find that 

\begin{equation}
\begin{split}
\mathcal{A}(h^-_x h^-_x h^+ ... h^+) = \ang{12}^4 \left(A_1\!\cdot \!\pd{P_{1}}\right) \left( A_2\!\cdot \!\pd{P_{2}} \right)\Bigg(\prod_{r \in R}\psi_{r,n}^{\ket{1}\ket{2}} \delta^4(P)\Bigg)
\end{split}
\end{equation}
where $A_{1,2}$ are vectors in the wavefunctions of particles 1,2.

\end{document}